\documentclass[prd,aps,showpacs,nofootinbib,floats,floatfix,superscriptaddress,twocolumn,letterpaper]{revtex4}

\usepackage{psfrag}
\usepackage[dvips]{graphicx}
\usepackage{amsmath}
\usepackage{amsfonts}
\usepackage{color}
\usepackage{dcolumn}
\usepackage{bm}
\usepackage{amsmath}

\newcommand{\hide}[1]{{}}

\def\laq{\raise 0.4ex\hbox{$<$}\kern -0.8em\lower 0.62ex\hbox{$\sim$}}
\def\gaq{\raise 0.4ex\hbox{$>$}\kern -0.7em\lower 0.62ex\hbox{$\sim$}}
\newcommand{\beq}{\begin{equation}}
\newcommand{\eeq}{\end{equation}}
\newcommand{\bea}{\begin{eqnarray}}
\newcommand{\eea}{\end{eqnarray}}
\newcommand{\ba}{\begin{array}}
\newcommand{\ea}{\end{array}}

\newcommand{\vphi}{\varphi} 
\newcommand{\pa}{\partial}

\newcommand{\vp}{\mbox{\boldmath${p}$}}

\begin{document}

\title{Toward faithful templates for non-spinning binary black holes using the effective-one-body approach}

\author{Alessandra Buonanno}

\author{Yi Pan} 

\affiliation{Maryland Center for Fundamental Physics, 
Department of Physics, University of Maryland, College Park, MD 20742}

\author{John G. Baker}

\author{Joan Centrella}

\author{Bernard J. Kelly}

\affiliation{Gravitational Astrophysics Laboratory, NASA 
Goddard Space Flight Center, 8800 Greenbelt Rd., Greenbelt, MD 20771}

\author{Sean T. McWilliams}

\affiliation{Department of Physics, University of Maryland, College Park, MD 20742}

\author{James R. van Meter}

\affiliation{Gravitational Astrophysics Laboratory, NASA 
Goddard Space Flight Center, 8800 Greenbelt Rd., Greenbelt, MD 20771}

\affiliation{Center for Space Science \& Technology, 
University of Maryland Baltimore
County, Physics Department, 1000 Hilltop Circle, Baltimore, MD 21250}

\begin{abstract}
We present an accurate approximation of the full 
gravitational radiation waveforms generated in the merger of non-eccentric
systems of two non-spinning black holes.
Utilizing information from recent numerical relativity simulations 
and the natural flexibility of the 
effective-one-body (EOB) model, we extend the latter so that it can successfully 
match the numerical relativity waveforms during the last stages of 
inspiral, merger and ringdown. By ``successfully'' here, we mean with phase differences 
$\laq\, 8 \%$ of a gravitational-wave cycle accumulated by the end of the 
ringdown phase, maximizing only over time of arrival and initial phase. 
We obtain this result by simply adding a $4$-post-Newtonian order 
correction in the EOB radial potential and determining the (constant) coefficient 
by imposing high-matching performances with numerical waveforms of  
mass ratios $m_1/m_2 = 1, 3/2, 2$ and $4$, $m_1$ and $m_2$ being the 
individual black-hole masses. The final black-hole mass and spin predicted 
by the numerical simulations are used to determine the ringdown frequency and decay time 
of three quasi-normal-mode damped sinusoids that are attached to the EOB inspiral-(plunge) waveform 
at the EOB light-ring. The EOB waveforms might be tested and further improved in the future by comparison with 
extremely long and accurate inspiral numerical-relativity waveforms. They may already be employed for coherent 
searches and parameter estimation of gravitational waves emitted by non-spinning coalescing 
binary black holes with ground-based laser-interferometer detectors.  
\end{abstract}
\date{\today}


\maketitle

\section{Introduction}
\label{sec1}

The network of ground-based laser-interferometer gravitational-wave (GW) 
detectors, such as LIGO~\cite{LIGO}, VIRGO~\cite{VIRGO}, GEO~\cite{GEO} 
and TAMA~\cite{TAMA}, are currently operating at design sensitivity 
(except for VIRGO which is expected to reach design sensitivity within one year) 
and are searching for GWs in the frequency range of $10 \mbox{--} 10^3$ 
Hz. Within the next decade these detectors will likely be complemented 
by the laser-interferometer space antenna (LISA)~\cite{LISA}, 
a joint venture between NASA and ESA, which will search for GWs 
in the frequency range $3 \times 10^{-5} \mbox{--} 10^{-1}$ Hz.

Binary systems composed of black holes (BHs) and/or neutron stars (NSs) 
are among the most promising GW sources. The search for GWs from 
coalescing binary systems and the extraction of parameters are based on the matched-filtering 
technique~\cite{MF}, which requires accurate knowledge of the waveform of the incoming signal. 
Recent comparisons~\cite{BCP,Goddshort,DA,Anatomy,NewJena,DNT,DNf} 
between numerical and post-Newtonian (PN) analytic 
waveforms emitted by non-spinning binary BH systems suggest that it 
should be possible to design purely analytic templates with the full numerics used to guide the
patching together of the inspiral and ringdown (RD) waveforms. 
This is an important avenue to template construction as
eventually thousands of waveform templates may be needed 
to extract the GW signal from the noise, an impossible 
demand for numerical relativity (NR) alone. 

The best-developed {\it analytic} method for describing the two-body dynamics 
of comparable-mass BHs and predicting the GW signal is undoubtedly the PN 
method~\cite{LB}, which for compact bodies is essentially an expansion in the 
characteristic orbital velocity $v/c$.  Template predictions are currently 
available through 3.5PN order ($v^7/c^7$)~\cite{JS98,DJSd,DJSPoincare,PNnospin},  
if the compact objects do not carry spin, and 2.5PN order ($v^5/c^5$)~\cite{PNspin} 
if they carry spin. Resummation of the PN expansion aimed 
at pushing analytic calculations until the final stage of evolution, 
including the transition inspiral--merger--ringdown,  have been 
proposed. In 1999, Buonanno \& Damour introduced
a non-perturbative resummation of the two-body conservative dynamics, 
the so-called effective-one-body (EOB) approach~\cite{BD1}.  
The original EOB model was computed using the 2PN conservative 
dynamics. It was then extended to 3PN order~\cite{DJS} when 
the 3PN calculation was completed~\cite{PNnospin} 
and then to spinning BHs~\cite{TD}. 
The EOB approach has been the {\it only} analytic approach able to predict, 
within $\sim 10\%$ of accuracy, the spin of the final BH~\cite{BD2}. 
Recently, by combining the EOB approach with test-mass limit 
predictions for the energy released during the merger-ringdown 
phases, Ref.~\cite{DNs} has refined this prediction 
obtaining $\sim 2\%$ of accuracy. 
The EOB approach also  provided a complete waveform, from inspiral to ringdown, 
for non-spinning~\cite{BD2,BD3} and spinning, precessing 
binary systems~\cite{BCD}. To include accurately 
the radiation-reaction contribution, the EOB approach uses the 
Pad\'e resummation of the GW flux, as proposed in Ref.~\cite{DIS}. 

By construction the EOB approach recovers exactly geodesic motion in 
the test-mass limit. In the comparable-mass limit the EOB approach provides 
a non-perturbative resummation of the dynamics, which today 
can be tested and improved by comparing it to NR results. NR simulations 
are in fact the best tool to describe the non-linear, strong-gravity 
regime of comparable-mass binary coalescences. 
As we shall see below, because of the reduction of the 
dynamics to {\it a few} crucial functions determining  
the inspiral evolution~\cite{BD1,BD2,TD}, and because of the rather 
simple procedure for matching the inspiral(-plunge) waveform 
to the ringdown waveform, the EOB model is particularly 
suitable for fitting to the numerical results
~\cite{first,second}. In this paper 
we shall employ its flexibility~\cite{DGG,TD} to obtain accurate  
waveforms for potentially the full range of 
non-spinning binary BHs. We shall test the analytic 
waveforms against the numerical ones for mass 
ratios ranging between $m_1/m_2=1 $ and $m_1/m_2 = 4$, with 
$m_1$ and $m_2$ being the BH masses. The method also allows us to 
predict the waveforms for mass ratios $m_1/m_2 > 4$.  
These waveforms will be tested against numerical results 
when accurate long numerical simulations 
for mass ratios $m_1/m_2 > 4$ become available.   
In this paper the comparisons are carried out using 
simulations from the NASA-Goddard group. 

The paper is organized as follows. In Sec.~\ref{sec2} we briefly review 
the EOB model. In Sec.~\ref{sec3} we {\it improve} the EOB model by 
adding a 4PN order unknown coefficient to the two-body conservative 
dynamics. In Sec.~\ref{sec4} we complete the EOB model using inputs 
from NR simulations.  In Sec.~\ref{secnew} we compare the improved EOB model 
to two accurate, long numerical simulations with 
mass ratios 1:1 and 4:1, determine the best-fit 4PN order 
coefficient and discuss the matching performances for several 
dominant modes. Section~\ref{sec5} summarizes 
our main conclusions. Appendix~\ref{appendix_massratio} refers to shorter numerical 
simulations with mass ratios 2:1 and 3:2.

\section{The effective-one-body model for non-spinning black hole binaries}
\label{sec2}

At the end of the 90s, in the absence of NR results and with the 
urgent need of providing templates to search for comparable-mass BHs, 
some resummation techniques of the post-Newtonian series were 
proposed. The general philosophy underlying these techniques~\cite{DIS,BD1}
 was to first {\it resum} in the test-mass--limit case the two crucial ingredients 
determining the gravitational-wave signal: the two-body energy and 
the gravitational-wave energy flux. In fact, in the test-mass--limit case  
these ingredients are known exactly. Secondly, it was assumed 
that the resummed quantities will {\it also} be a good 
representation of the comparable-mass case, viewed  
as a smooth deformation of the test-mass--limit case.

The resummation technique discussed in this section, 
the EOB approach~\cite{BD1}, was originally inspired by a similar approach 
introduced by Br\'ezin, Itzykson and Zinn-Justin~\cite{BIZ} to study 
two electromagnetically interacting particles with comparable masses. 
The basic idea of the EOB approach is to map the {\em real} 
conservative two-body dynamics up to the highest PN order available, 
onto an {\em effective} one-body problem, 
where a test particle of mass $\mu=m_1 m_2/M$, with  
$m_1$, $m_2$ the BH masses and $M=m_1+m_2$, moves
in some effective background metric $g_{\mu \nu}^{\rm eff}$. 
This mapping has been worked out within the Hamilton-Jacobi formalism, 
by requiring that while the action variables of the real 
and effective description 
coincide (i.e.\ ${L_{\rm real}} = {L_{\rm eff}}$, ${{\cal I}_{\rm real}}= {{\cal I}_{\rm eff}}$, 
where $L$ denotes the total angular momentum, and 
${\cal I}$ the radial action variable), the energy axis is allowed to change: 
${{E}_{\rm real}} = f({{E}_{\rm eff}})$, where $f$ is a generic function 
determined by the mapping. 
By applying the above rules defining the mapping, it was 
found~\cite{BD1} in the non-spinning case that as long as radiation-reaction effects are not taken 
into account, the effective metric is just a deformation 
of the Schwarzschild metric, with deformation parameter 
$\eta = \mu/M$. 

The explicit expression of the non-spinning EOB effective Hamiltonian 
through 3PN order is~\cite{BD1,DJS}:
\begin{widetext}
\bea 
\label{eq:genexp}
H_{\rm eff}(\mathbf{r},\vp) = \mu\, 
\widehat{H}_{\rm eff}({\mathbf r},{\mathbf p}) 
= \mu\,\sqrt{A (r) \left[ 1 + 
{\mathbf p}^{2} +
\left( \frac{A(r)}{D(r)} - 1 \right) ({\mathbf n} \cdot {\mathbf p})^2
+ \frac{1}{r^{2}} \left( z_1 ({\mathbf p}^{2})^2 + z_2 \, {\mathbf p}^{2}
({\mathbf n} \cdot {\mathbf p})^2 + z_3 ({\mathbf n} \cdot {\mathbf p})^4 \right) \right]} \,,
\eea
\end{widetext}
with $\mathbf{r}$ and $\mathbf{p}$ being the reduced dimensionless variables; 
$\mathbf{n} = \mathbf{r}/r$ where we set $r = |{\mathbf{r}}|$. In the absence of 
spins the motion is constrained to a plane. Introducing polar coordinates 
$(r,\varphi, p_r,p_\varphi)$, the EOB effective metric reads
\begin{widetext}
\beq 
ds_{\rm eff}^2 \equiv g_{\mu \nu}^{\rm eff}\,dx^\mu\, dx^\nu =
-A(r)\,c^2dt^2 + \frac{D(r)}{A(r)}\,dr^2+
r^2\,(d\theta^2+\sin^2\theta\,d\varphi^2) \,.  
\eeq
\end{widetext}
The EOB real Hamiltonian is 
\beq
\label{himpr}
H_{\rm real} =  M\,\sqrt{1 + 2\eta\,\left ( \frac{H_{\rm eff} -\mu}{\mu}\right )} -M\,,
\eeq
and we define $\hat{H}_{\rm real} = H_{\rm real}/\mu$.  
Remarkably, as originally observed in Ref.~\cite{BD1}, the mapping between the real and the effective 
Hamiltonians given by Eq.~(\ref{himpr}) coincides with the mapping
obtained in the context of quantum electrodynamics in Ref.~\cite{BIZ}, where 
the authors mapped the one-body relativistic Balmer formula onto the two-body 
energy formula. Moreover, Eq.~(\ref{himpr}) holds at 2PN and 3PN order~\cite{DJS}.   
The coefficients $z_1,z_2$ and $z_3$ in Eq.~(\ref{eq:genexp}) 
are arbitrary, subject to the constraint
\beq 8z_1 + 4z_2 +3z_3 = 6(4-3\eta)\,\eta\,.  
\eeq
The coefficients $A(r)$ and $D(r)$ in Eq.~(\ref{eq:genexp}) have been calculated through 
3PN order~\cite{BD1,DJS}. In Taylor-expanded form they read:
\beq
\label{coeffA}
A_{T}^{\rm 3PN}(r) = 1 - \frac{2}{r}+\frac{2\eta}{r^{3}}+ \left [ \left (
\frac{94}{3}-\frac{41}{32}\pi^2\right )\,\eta -z_1\right ]\,\frac{1}{r^{4}}\,,
\eeq
\beq
\label{coeffD}
D_{T}^{\rm 3PN}(r) = 1 -\frac{6\eta}{r^{2}}+\left [7{z}_1 + {z}_2+ 2\eta\, (3\eta-26)\right ]\,
\frac{1}{r^{3}}\,. 
\eeq
In principle we could explore the possibility of determining some of 
the $z_i$ coefficients through a fit with the numerical results. 
However, here we do not follow this possibility and, as in previous 
works, except for Ref.~\cite{bcv1}, we set ${z}_1={z}_2=0$, 
$z_3 = 2(4 - 3\eta)\eta$. The EOB effective potential $A_{T}^{\rm 3PN}(r)$ does not lead to 
a last-stable circular orbit (LSO), contrary to what happens in the 2PN-accurate case \cite{BD1}. 
This is due to the rather large value of the 3PN coefficient 
${94}/{3}-{41}/{32}\pi^2 \simeq 18.688$ entering the PN expansion
of $A(r)$. Replacing the PN-expanded form of $A(r)$ 
by a Pad\'e approximant cures this problem~\cite{DJS}. The Pad\'e approximant is
\beq
A_{P_2^1}^{\rm 2PN}(r) = \frac{r\,(-4+2r+\eta)}{2r^{2}+2\eta+r\,\eta}\,,
\label{coeffPA2}
\eeq
at 2PN order and
\begin{widetext}
\beq
\label{coeffPA}
A_{P_3^1}^{\rm 3PN}(r) = \frac{r^{2}\,[(a_4(\eta,0)+8\eta-16) + r\,(8-2\eta)]}{
r^{3}\,(8-2\eta)+r^{2}\,[a_4(\eta,0)+4\eta]+r\,[2a_4(\eta,0)+8\eta]
+4[\eta^2+a_4(\eta,0)]}\,,
\eeq
\end{widetext}
at 3PN order where
\beq \label{a4}
a_4(\eta,z_1) = \left [\left (\frac{94}{3}-\frac{41}{32}\pi^2 \right )\,\eta - {z}_1\right ]\,.
\eeq
For the coefficient $D(r)$ at 3PN order we use the Pad\'e approximant 
\beq
\label{D3PN}
D_{P_3^0}^{\rm 3PN}(r)=\frac{r^3}{r^3+6\eta r+2 \eta(26-3\eta)}\,.
\eeq
To include radiation-reaction effects we write the EOB Hamilton equations 
in terms of the reduced quantities $\widehat{H}$, $\widehat{t} 
= t/M$, $\widehat{\omega} = \omega\,M$~\cite{BD2}, as
~\footnote{When using the EOB real Hamiltonian we should in principle
consider the (generalized) canonical transformation between the real and 
effective variables which is explicitly given as a PN expansion in 
Refs.~\cite{BD1,DJS}. However, since the Hamilton equations are valid in 
any canonical coordinate system, when we evolve the EOB dynamics we 
write the Hamilton equations in terms of the effective variables. When comparing 
to NR results, there might be some differences in the time variable, though. 
In any case Eqs.~(\ref{eq:eobhamone})--(\ref{eq:eobhamfour}) define our EOB model.}  
\begin{figure*}
\includegraphics[width=0.35\textwidth]{LRa5.eps} \hspace{1cm}
\includegraphics[width=0.35\textwidth]{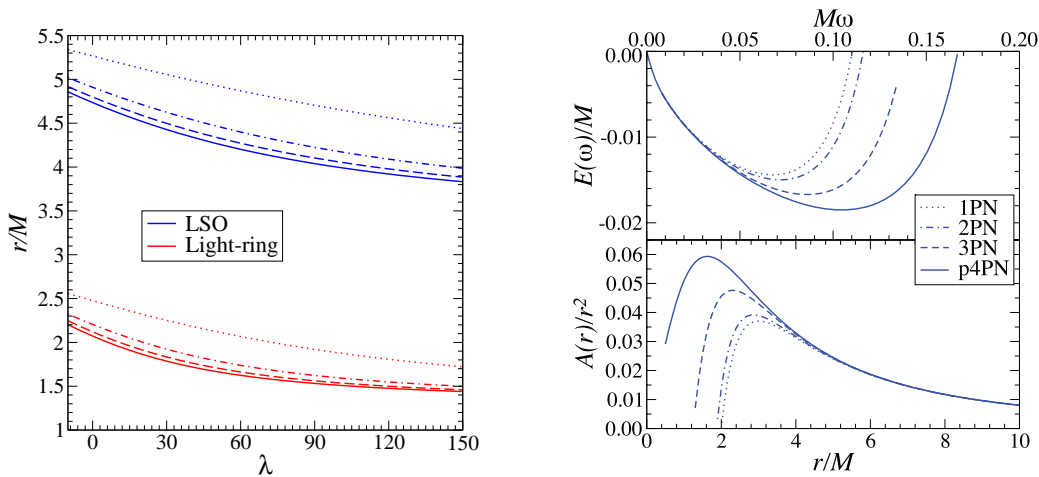} 
\caption{In the left panel we show the position of the LSO and light-ring as function 
of the parameter $\lambda$, for different mass ratios: 4:1 (dotted line), 
2:1 (dot-dashed line), 3:2 (dashed line) and 1:1 (continuous line). 
In the right panel we show: (top part) the energy for circular 
orbits as a function of the frequency 
evaluated from the EOB Hamiltonian, (bottom part) the radial potential 
as function of the radial coordinate for a massless particle in the EOB model. 
The various curves refer to different PN orders. 
\label{fig1}}
\end{figure*}
\bea 
\frac{dr}{d \widehat{t}} &=& \frac{\pa \widehat{H}}{\pa p_r}(r,p_r,p_\vphi)\,, \label{eq:eobhamone} \\
\frac{d \vphi}{d \widehat{t}} &\equiv& \widehat{\omega} = \frac{\pa \widehat{H}}
{\pa p_\vphi}(r,p_r,p_\vphi)\,, \label{eq:eobhamtwo} \\
\frac{d p_r}{d \widehat{t}} &=& - \frac{\pa \widehat{H}}{\pa r}(r,p_r,p_\vphi)\,, \\
\frac{d p_\vphi}{d \widehat{t}} &=&
\widehat{F}^\vphi[\widehat{\omega} (r,p_r,p_{\varphi})]\,, 
\label{eq:eobhamfour}
\eea
where for the $\varphi$ component of the radiation-reaction force we
shall use the P-approximant~\cite{DIS,BD2} 
\beq
\label{fluxP}
\widehat{F}_{P_N}^\vphi \equiv 
- \frac{1}{\eta\,v_\omega^3}\,
{\cal F}_{P_N}[v_\omega]=
- \frac{32}{5}\,\eta\,v_\omega^7\,
\frac{f_{P_N}(v_\omega;\eta)}{1 - v_\omega/v_{\rm pole}(\eta)}\,,
\eeq
where $v_\omega \equiv \widehat{\omega}^{1/3} \equiv (d \varphi/d\widehat{t})^{1/3}$. 
The coefficients $f_{P_N}$ can be read from Eqs.~(50)--(54) in Ref.~\cite{bcv1}, 
while for $v_{\rm pole}$ we use the expression given by Eq.~(55) in 
Ref.~\cite{bcv1}. Initial conditions for Eqs.~(\ref{eq:eobhamone})--(\ref{eq:eobhamfour}) are 
discussed in Ref.~\cite{BD2}. In Ref.~\cite{DG,DNT}, the authors pointed out that 
in principle a more accurate expression of the radiation-reaction force should 
not use the Keplerian relation between $r$ and $\omega$ when the binary evolves inside the LSO. 
However, as also noticed and discussed in Ref.~\cite{DNf}, this modification of the radiation-reaction effects 
has little effect on the waveform amplitude. Since it is not one of the goals 
of this paper to improve the amplitude agreement between the numerical and 
EOB waveforms, we do not include it.

The last crucial ingredient of the EOB model is the inclusion of the 
ringdown phase. After the two BHs merge, the system settles down to a Kerr BH 
and emits quasi-normal modes (QNMs)~\cite{qnm,Press}. 
In the test-mass limit, $\eta \ll 1$, Refs.~\cite{Davis,Press} realized 
that when a test particle falls radially below $ 3 M$ (the 
unstable light-ring of Schwarzschild), it immediately triggers 
the production of QNMs, thus producing a universal merger signal 
(by contrast the direct gravitational radiation from the source is 
strongly filtered by the curvature potential 
barrier centered around it, see Fig.~\ref{fig7}). 
In the comparable-mass case $\eta \,\laq \, 1/4$, to approximate 
the late part of the merger waveforms, 
Ref.~\cite{CLA}  proposed the so-called close-limit approximation, 
which consists of switching from the two-body description to the 
one-body description (perturbed-BH) close to the light-ring 
location. Based on these observations, Ref.~\cite{BD1} modeled 
the merger as a very short (instantaneous) phase 
and matched the inspiral(-plunge) waveform to a damped sinusoid 
at the light-ring position. The frequency and decay time 
were computed estimating the final BH mass and spin 
from the EOB energy and angular momentum 
at the matching point. The matching procedure has then  
been improved, by adding more QNMs, extending it 
to several multipole moments~\cite{DG,BCP}, and 
applying it over a time-interval instead of 
one point in time~\cite{DNf}. 

\section{The pseudo 4PN  effective-one-body model}
\label{sec3}

Previous investigations~\cite{BCP,DA} focusing on comparable-mass binaries, have shown 
that a non-negligible dephasing can accumulate at the transition inspiral(-plunge) 
to ringdown between the 3.5PN-EOB waveform and the NR 
waveform. The dephasing is caused by the much faster increase of the GW frequency in the 
3.5PN-EOB model than in the NR simulation when approaching the light-ring position.   
Although the dephasing would prevent an accurate determination of the binary parameters
in applications to gravitational wave observations, 
it would not prevent detection of the signals. In Ref.~\cite{DA} the authors 
built {\it effectual}~\footnote{Following Ref.~\cite{DIS}, by {\it effectual} templates we mean 
templates that have large overlaps, say $ \gaq \, 96.5\%$, with the 
expected signal after maximizing over the initial phase, time of arrival 
and BH masses. Effectual templates can be used for detection 
but may lead to large biased in estimating the binary parameters.} 
EOB templates which match the EOB inspiral(-plunge) 
to ringdown through three parameters. The latter describe the time of matching, and 
the difference between the final BH mass (spin) and the EOB energy (angular momentum) 
at the matching point~\footnote{These parameters are not arbitrary. They 
can be related to the BH masses as long as accurate NR simulations for several 
BH mass ratios are available.}.  In this paper we improve on Ref.~\cite{DA} 
implementing a matching procedure that does not require the introduction 
of any new parameter and 
that agrees with numerical simulation waveforms within 
a rather small phase difference, $\laq \, 0.08 $ of a GW cycles, 
thus providing accurate or {\it faithful}~\footnote{Following Ref.~\cite{DIS},
by {\it faithful} templates we mean templates that have
large overlaps, say $ \gaq \, 96.5\%$, with the 
expected signal maximizing {\it only}
 over the initial phase and time of arrival.} templates. 

\begin{table*}
\begin{center}
{
\begin{tabular}{|l|c|c|c|c|c|c|c|}\hline
$\eta$ & \multicolumn{1}{c|}{$[M_f/M]_{\rm Goddard}$} &
\multicolumn{1}{c|}{$[a_f/M_f]_{\rm Goddard}$} 
& \multicolumn{1}{c|}{$[M_f/M]_{\rm fit}$} &
\multicolumn{1}{c|}{$[a_f/M_f]_{\rm fit}$} &
\multicolumn{1}{c|}{$[M_f/M_o]_{\rm Jena}$} &
\multicolumn{1}{c|}{$[a_f/M_f]_{\rm Jena}$} 
\\ \hline\hline 
$0.25$ & 0.9526 & 0.687 & 0.9546 & 0.685 & 0.9628 & 0.684\\
$0.24$ & 0.9561 & 0.670 & 0.9576 & 0.664 & 0.9660 & 0.664 \\
$0.2222$ & 0.9668 & 0.621 & 0.9633 & 0.623 & 0.9714 & 0.626 \\
$0.2041$ & $\cdots$  & $\cdots$  & 0.9676 & 0.586 & 0.9762 & 0.581 \\
$0.1875$ & $\cdots$  & $\cdots$  & 0.9618 & 0.548 & 0.9800 & 0.544 \\
$0.1728$ & $\cdots$  & $\cdots$  & 0.9752 & 0.512 & 0.9831 & 0.509 \\
$0.16$ & 0.9783 & 0.472 & 0.9781 & 0.480 & 0.9855 & 0.478 \\ 
$0.12$ & $\cdots$ & $\cdots$ & 0.9860 & 0.374 & $\cdots$ & $\cdots$\\
$0.08$ & $\cdots$ & $\cdots$ & 0.9922 & 0.259 & $\cdots$ & $\cdots$ \\ 
$0.04$ & $\cdots$ & $\cdots$ & 0.9969 & 0.134 & $\cdots$ & $\cdots$ \\
$0.01$ & $\cdots$ & $\cdots$ & 0.9994 & 0.034 & $\cdots$ & $\cdots$ \\ 
\hline
\end{tabular}}\end{center}
\caption{For several values of $\eta$, we list in the second and third columns 
the values of $M_f/M$ and $a_f/M_f$ computed from the energy released and by extracting the fundamental 
QNM from $_{-\!2}C_{22}$, respectively. In the fourth and fifth columns 
we list the values obtained using the one-parameter fits 
$M_f/M = 1 + (\sqrt{8/9}-1)\,\eta -0.498 (\pm 0.027)\,\eta^2$, and $a_f/M_f = \sqrt{12}\,\eta -2.900 (\pm 0.065)\,\eta^2$,
where the terms linear in $\eta$ have been fixed to the test-mass limit values. In the 
last two columns we list the values from the Jena group (see Table V of Ref.~\cite{berti07}).  Note that the Jena mass values are scaled differently, against 
ADM mass, $M_o$ of the system at the beginning of the numerical simulations, 
which should result slightly larger values.
\label{tab1}}
\end{table*}

To decrease the differences between the EOB and NR waveforms during the last stages 
of inspiral and plunge, we introduce a 4PN order term in the effective 
potential $A(r)$, given by Eq.~(\ref{coeffA}), that is 
\beq
A_T^{\rm p4PN}(r) = A_T^{\rm 3PN}(r) + \frac{a_5(\eta)}{r^5}\,, 
\eeq
and Pad\'e-approximate it using the approximant $A_{P_4^1}$. 
A similar modification was employed in Ref.~\cite{DGG} 
to obtain better matches of the EOB model to quasi-equilibrium 
initial-data configurations~\cite{GGB} and was also pointed out 
in Ref.~\cite{TD}. An interesting motivation 
for this change is the following~\cite{BD1,BD2}: From 
Eq.~(\ref{eq:eobhamtwo}) it is straightforward to write the 
EOB {\it instantaneous} frequency as
\beq
\omega(t) = \frac{A(r)}{r^2}\,\frac{p_\varphi}{\eta \hat{H}_{\rm real} \hat{H}_{\rm eff}}\,.
\eeq
It is reasonable to assume that during the plunge, the two-body dynamics 
is no longer driven by radiation-reaction effects~\cite{BD3}, but occurs mostly 
along a geodesic, with fixed angular-momentum $p_\varphi$ and energy 
$\hat{H}_{\rm real}$ and $\hat{H}_{\rm eff}$. Thus, we can write 
\bea
\omega_{\rm plunge}(t) &=& \frac{A(r)}{r^2}\,{\rm const}\,, \\
{\rm const} &=& \left [\frac{p_\varphi}{\eta \hat{H}_{\rm real} \hat{H}_{\rm eff}} \right ]_{\rm LSO}\,.
\label{omegap}
\eea
The above Eq.~(\ref{omegap}) clearly shows how the coefficient $A(r)$ determines the 
frequency during the plunge, i.e., from the LSO until the light-ring position. 
The latter happens at the maximum of $A(r)/r^2$. 
A direct test has shown that the relative difference between $\omega_{\rm plunge}$ and 
the exact $\omega$ from the EOB-LSO to the light-ring is at most $5\%$ in the 
case $4:1$.  

Since the 4PN term has not been calculated in PN theory, we shall denote 
it as ``p4PN'', where ``p'' stands for {\it pseudo}. We do 
not claim that when the 4PN order term is calculated it will 
agree with the p4PN order term. In fact, the latter should be considered 
as a phenomenological term. We have
\beq
A_{P_4^1}^{\rm p4PN}(r) = \frac{{\rm Num}(A_{P_4^1}^{\rm p4PN})}{{\rm Den}(A_{P_4^1}^{\rm p4PN})}\,,
\eeq
with 
\begin{widetext}
\beq
{\rm Num}(A_{P_4^1}^{\rm p4PN}) = r^3\,[32 -24 \eta - 
4 a_4(\eta,0) - a_5(\eta,\lambda)] + r^4 [a_4(\eta,0)  - 16 +8 \eta]\,,
\eeq
\bea
{\rm Den}(A_{P_4^1}^{\rm p4PN}) &=& -a_4^2(\eta,0)  - 8a_5(\eta,\lambda) - 
8 a_4(\eta,0)  \eta + 2 a_5(\eta,\lambda) \eta - 16 \eta^2 + r\,[-8 a_4(\eta,0)  -4 a_5(\eta,\lambda)  
-2 a_4(\eta,0)  \eta - \nonumber \\ 
&& 16 \eta^2] + r^2\,[-4a_4(\eta,0)  - 2a_5(\eta,\lambda) -16 \eta]+r^3\,[-2 a_4(\eta,0)  -a_5(\eta,\lambda) - 8 \eta] 
+ r^4\, (-16 + a_4(\eta,0)  + 8 \eta)\,, \nonumber \\
\eea
\end{widetext}
\begin{figure*}
\includegraphics[width=0.35\textwidth]{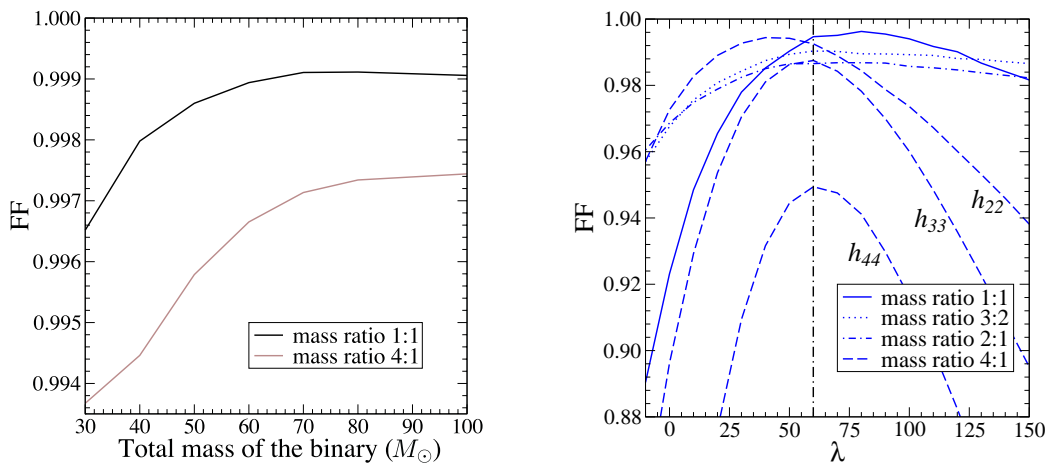} \hspace{1cm}
\includegraphics[width=0.35\textwidth]{FFa5.eps}  
\caption{In the left panel we show the (minmax) FF between 
the high and medium resolution runs, $\langle h^{\rm NR,h}, 
h^{\rm NR,m} \rangle $, as a function of the binary
total-mass $M$. The ${\rm FF}$s are evaluated using LIGO's PSD. 
If we use white noise we find 0.9922 and 0.9920 for mass ratios $1:1$ and 
$4:1$,  respectively. In the right panel, for different mass ratios, 
we show how the (minmax) FF $\langle h^{\rm NR}, h^{\rm EOB} \rangle $ 
(computed using white noise) depends on the parameter $\lambda$. 
For mass ratios $1:1$, $2:1$, and $3:2$ we compute 
$\langle h_{22}^{\rm NR}, h_{22}^{\rm EOB} \rangle $, while for 
$4:1$ we show also results for $\langle h_{33}^{\rm NR}, h_{33}^{\rm EOB} \rangle $ 
and $\langle h_{44}^{\rm NR}, h_{44}^{\rm EOB} \rangle $.
The vertical line refers to the value $\lambda=60$ which we employ in 
all subsequent analyses. \label{fig4}}
\end{figure*}
where
\beq \label{a5}
a_5(\eta,\lambda) = \lambda\,\eta\,,
\eeq
and $\lambda$ will be determined by comparison with numerical results. 
We could also introduce a 4PN order term in the coefficient $D(r)$. However, we find 
that the effect on the dynamics is relatively small and decide to use the Pad\'e approximant 
\beq
D_{P_4^0}^{\rm p4PN}(r)=\frac{r^4}{r^4+6\eta r^2+2\eta(26-3\eta)r+36\eta}\,.
\eeq
The difference between $D_{P_4^0}^{\rm p4PN}(r)$ and $D_{P_3^0}^{\rm 3PN}(r)$, Eq.~(\ref{D3PN}), causes a
negligible change in all our results. 

In the left panel of Fig.~\ref{fig1} we show how the p4PN 
order term $\lambda$ modifies the position of 
the EOB LSO and light-ring (the last unstable orbit for a massless 
particle) for several binary mass ratios. Later on we shall see that 
the value of $\lambda$ that best fits the NR results (see Sec.~\ref{sec4}) 
is $\lambda=60$. It always guarantees the presence of a LSO and a light ring.
In the right panel of Fig~\ref{fig1}, we show the circular-orbit 
energy computed with the EOB Hamiltonian, and the radial potential 
for a massless particle, at different PN orders 
with fixed $\lambda = 60$. We notice that the LSO energy ($E^{\rm EOB}_{\rm p4PN}/M 
= - 0.0185$) and frequency ($M \omega^{\rm EOB}_{\rm p4PN} = 0.1047$) at 
p4PN order are closer to the corresponding values obtained using 
the 3PN-Taylor-expanded model for quasi-circular adiabatic orbits~\cite{ICO}, 
and to the quasi-equilibrium initial-data approach~\cite{CP} 
(see Fig. 16 and Table II
in Ref.~\cite{CP}). This could be a pure accident. 
In fact, it should be kept in mind that the LSO frequency computed 
from the 3PN-Taylor-expanded conservative dynamics is~\cite{ICO} 
$M \omega_{\rm 3PN}^{\rm T} \sim 0.129$ ($E^{\rm T}_{\rm 3PN}/M = - 0.0193$), 
quite close to the formation of the common apparent horizon in the NR simulation, and quite far from 
the frequency $\sim 0.08$ at which the indistinct plunge occurs~\cite{BCP}.
What we can certainly say is that the p4PN-EOB conservative dynamics is 
closer than at 3PN order to the 3PN Taylor-expanded conservative dynamics 
of quasi-circular adiabatic orbits~\cite{ICO}. 

\begin{figure*}
\includegraphics[width=0.35\textwidth]{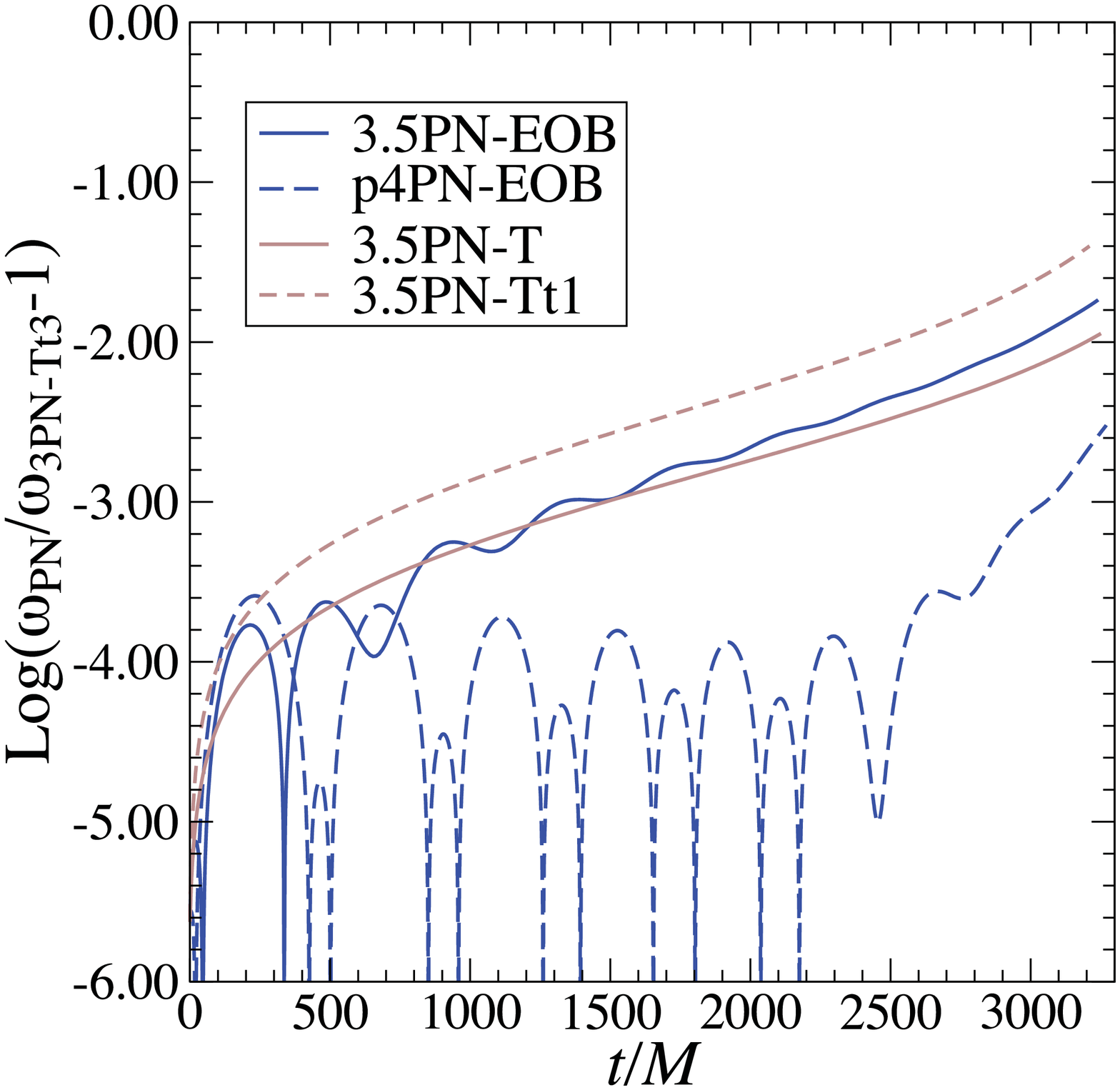} \hspace{1cm}
\includegraphics[width=0.35\textwidth]{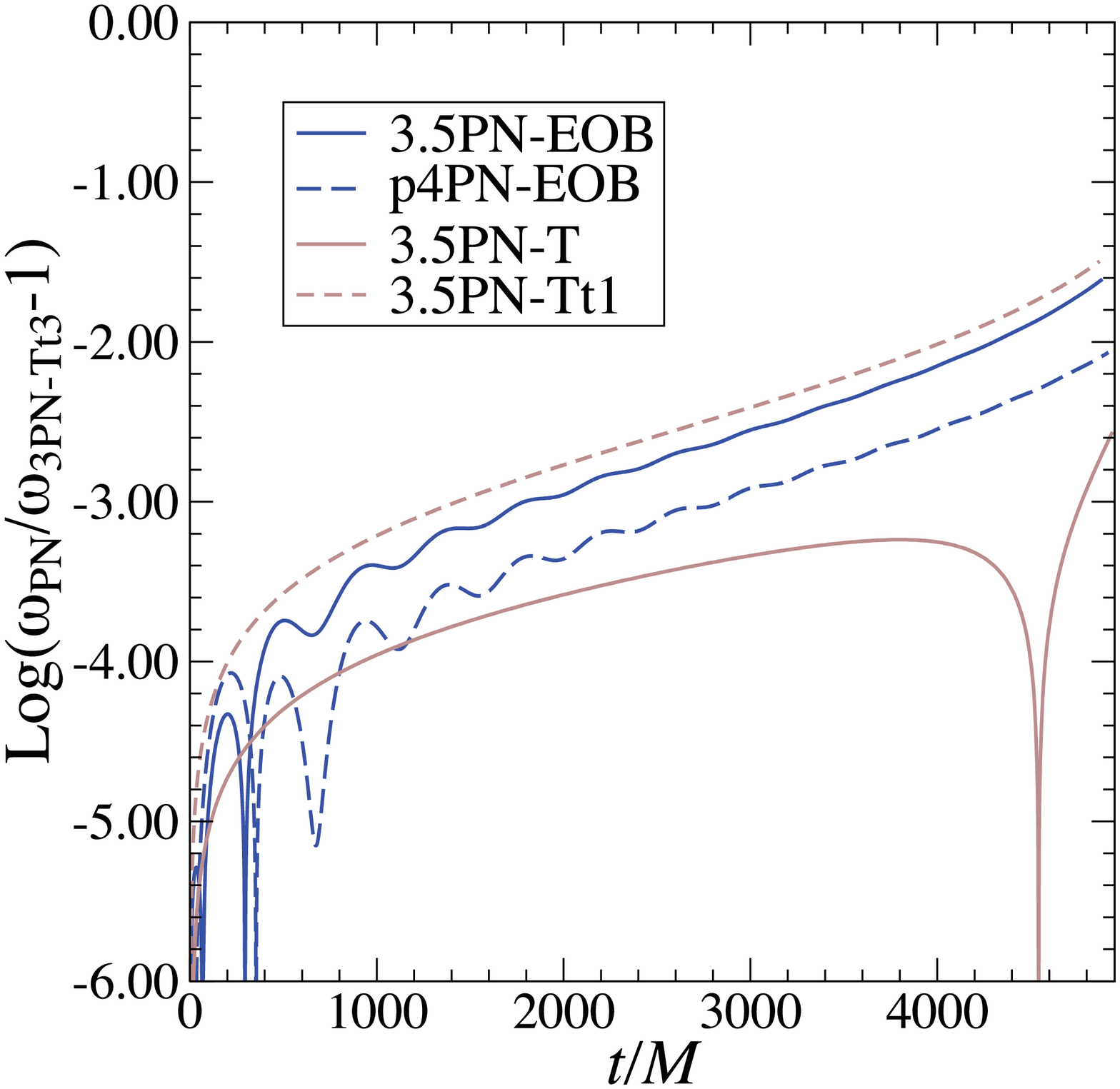} 
\caption{We show the differences in the orbital frequency between the 3.5PN-Tt3 model 
and 3.5PN-T, 3.5PN-Tt1, 3.5PN-EOB, p4PN-EOB models for mass ratios 1:1 (left panel) and 4:1 (right panel). 
The PN frequencies coincide at $\omega M = 0.017$ at $t=0$, and end at $\omega M = 0.035$. 
\label{figfreq}}
\end{figure*}
\section{Effective-one-body waveforms for inspiral, merger and ringdown}
\label{sec4}

Integrating the p4PN-EOB Hamiltonian equations provides a description of the binary's dynamical evolution.
As described below, we derive our waveforms directly from the EOB dynamics
until the system approaches the light-ring.  Thereafter we complete each 
spherical harmonic waveform component by matching it to a set of quasinormal
ringdown modes.  Because the ringdown mode frequencies depend on the 
mass $M_f$ and spin parameter $a_f$ of the final BH formed by 
the merger, this part of the model will require an additional prescription for 
accurately determining these values.

Following the tradition in NR we describe the waveforms in
terms of a spin-weighted spherical harmonic decomposition. 
From our numerical simulations we directly compute the Weyl tensor $\Psi_4$, 
which in terms of spin-weight $-2$ spherical harmonics $_{-2}Y_{l
m}(\theta,\phi)$~\cite{wiaux} reads [see Ref.~\cite{Goddlong} for details]
\beq
  M\, R\,\Psi_4 = 
\sum_{l m}{}_{-\!2}C_{l m}(t){}_{-\!2}Y_{l m}(\theta,\phi)\,,
\eeq
$R$ being the extraction radius. In terms of the $+$ and $\times$ GW polarizations 
we have 
\footnote{Note that this definition of $\Psi_4$  is tetrad-dependent.  Here we assume 
the tetrad given in Ref.~\cite{BCL}, Eqs.(5.6).
It is also common for $\Psi_4$ to be scaled according
to an asymptotically Kinnersley tetrad (Ref.~\cite{BCL}, Eqs.(5.9)) which introduces a factor of 2 as
in Ref.~\cite{Goddlong}}.
\beq
\Psi_4 = -(\ddot{h}_+ - i\ddot{h}_\times)\,.
\eeq
Thus, we can write
\beq
{}_{-\!2}C_{lm} = - M\,R\,\int d \Omega\, {}_{-\!2}Y^*_{l m}(\theta,\phi)\,(\ddot{h}_+ - i\ddot{h}_\times)\,.
\eeq
In the adiabatic approximation ($\dot{\omega}/\omega^2 \ll 1$), we obtain 
\beq
{}_{-\!2}C_{l m} = -m^2 \,\omega^2\,h_{lm}\,, 
\eeq
where 
\beq
h_{lm } \equiv -(h_+ - i h_\times)_{lm} = -\int d \Omega\, {}_{-\!2}Y^*_{l m}(\theta,\phi)\,({h}_+ - i{h}_\times)\,.
\eeq
We compute the EOB $h_{lm}$ in the so-called restricted approximation, i.e., 
at leading order in the PN expansion. They read:
\bea
\label{22}
h^{\rm EOB}_{21}&=& -\frac{8}{3}\,\sqrt{\frac{\pi}{5}}\,\frac{\delta m}{M}\,\eta \,(M\,\omega)\,\,e^{-i\,\varphi}\,,
 \\
h^{\rm EOB}_{22}&=& -8\,\sqrt{\frac{\pi}{5}}\,\eta\,(M\,\omega)^{2/3}\,\,e^{-2i\,\varphi}\,, \\
h^{\rm EOB}_{31} &=&- \frac{2}{3}\,\sqrt{\frac{\pi}{70}}\,\eta\,\frac{\delta m}{M}
\,(M\,\omega)\,e^{-i\,\varphi}\,,\\
h^{\rm EOB}_{32} &=&- \frac{8}{3}\,\sqrt{\frac{\pi}{7}}\,\eta\,(1-3\,\eta)\,(M\,\omega)^{4/3}\,e^{-2i\,\varphi}\,,\\
h^{\rm EOB}_{33} &=& -3\,\sqrt{\frac{6\pi}{7}}\,\frac{\delta m}{M}\,\eta\,(M\,\omega)\,e^{-3i\,\varphi}\,,\\
h^{\rm EOB}_{42} &=& -\frac{8}{63}\,\sqrt{\pi}\,\eta\,(1-3\eta)\,(M\,\omega)^{4/3}\,e^{-2i\,\varphi}\,,\\
h^{\rm EOB}_{44} &=& -\frac{64}{9}\,\sqrt{\frac{\pi}{7}}\,\eta\,(1-3\eta)\,(M\,\omega)^{4/3}\,e^{-4i\,\varphi}\,,
\label{44}
\eea
where $\delta m = m_1-m_2>0$, and $\varphi$ is the binary orbital phase. Note that $h_{l -m} = (-1)^l\,h_{lm}^*$.
The RD modes are attached at the time when the orbital frequency reaches its maximum
and this occurs slightly before the light-ring position, 
$r_{\rm match} = 1.651$, $M \omega = 0.1883$ ($\eta=0.25$) and 
$r_{\rm match} = 2.089$, $M \omega = 0.1665$ ($\eta = 0.16$). It is useful to have an analytic 
formula relating the maximum of the orbital frequency, i.e., 
the matching point, to $\eta$. A simple fit gives
\begin{figure*}
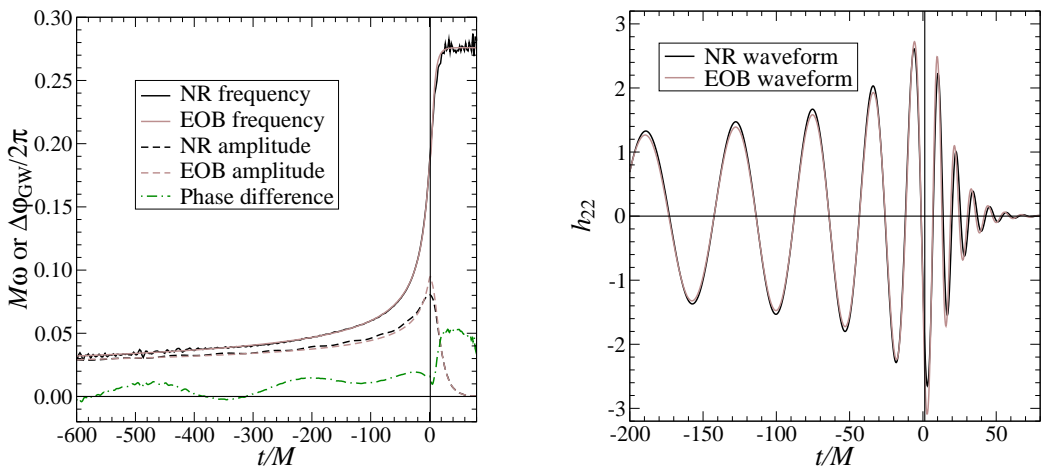

\begin{center}
\includegraphics[width=0.35\textwidth]{others5050.eps}  \hspace{1cm}
\includegraphics[width=0.35\textwidth]{waves5050.eps} 
\caption{Equal-mass binary. In the left panel we plot the NR and p4PN-EOB frequencies 
and amplitudes, and the phase difference between the EOB and NR $h_{22}$. 
In the right panel we compare the EOB and NR $h_{22}$. 
We maximize only on the initial phase and time of arrival. Note that 
we show only the last few cycles. The complete inspiral waveform has 14 GW cycles. 
\label{fig2}}
\end{center}
\end{figure*}
\beq
\label{omegamatch}
M\,\omega_{\rm match} = 0.133 + 0.183\,\eta + 0.161 \eta^2\,. 
\eeq
We find that the above fitting has $ < 0.35 \%$ error comparing to numerical 
values in the range $\eta = 0.05 \mbox{--} 0.25$.  
The matching to QNMs is obtained by imposing the continuity of $h_{l m}(t)$ 
and all the higher time derivatives needed to fix the six unknown amplitudes 
and phases of the three RD modes~\cite{BD2,BCP}. Following Ref.~\cite{BCP}, 
we attach the fundamental mode, and two overtones. We find that 
the matching-performance does not improve significantly if we 
add more overtones. The frequency and decay time of the RD modes 
are computed using the mass $M_f$ and spin  $a_f$ of the final BH, using 
Refs.~\cite{RD}.

For non-spinning binary systems, it is now possible to determine how the final-BH mass 
and spin depend on the mass ratio. Here we apply an empirical estimate for the functions $M_f(\eta)$ as
and $a_f(\eta)$ based on a combination of numerical simulations
in the range $\eta>0.16$ and expectations from the test particle limit
$\eta\rightarrow 0$.  In Table~\ref{tab1} we list the final BH masses and spins 
for $\eta = 0.25, 0.24, 0.22, 0.16$, extracted from the NR simulations. 
The values are compatible with Ref.~\cite{recoil,berti07}. 
The final mass was computed from the difference of the total radiated energy
and the initial ADM mass.  To obtain the final spin of the merged BH we first 
calculate the complex QNM frequency of the $l=2$, $m=2$ mode by a linear 
fit to the phase and log amplitude; the two slopes give the real
and imaginary parts of the frequency which are uniquely related to 
the final spin~\cite{RD}.

In the absence of NR results, to determine the final BH masses and spins to lower 
values of $\eta$ we apply a fit  to the data $\eta = 0.25, 0.24, 0.22, 0.16$. 
For the BH mass we consider the one-parameter fit function $M_f/M = 1 + (\sqrt{8/9}-1) \eta 
-0.498 (\pm 0.027)\, \eta^2$, where 
the coefficient of the linear term in $\eta$ has been fixed to the test-mass limit value.  
For the BH spin we employ the one-parameter fit function $a_f/M_f =  \sqrt{12} \eta 
-2.900 (\pm 0.065)\, \eta^2$, where again 
the linear term in $\eta$ has been fixed to the test-mass limit prediction. 
If we were using a two-parameter fit  we would obtain: 
$M_f/M = 1 -0.024 (\pm 0.057) \,\eta -0.641 (\pm 0.249)\,\eta^2$, 
and $a_f/M_f = 3.29 (\pm 0.08) \,\eta -2.13 (\pm 0.33)\,\eta^2$. 
We notice that the value of $3.29$ is quite close to the LSO angular-momentum for 
a test particle in Schwarzschild, i.e.,  $\sqrt{12} \simeq 3.4641$~\cite{berti07}. 
However, since the two-parameter fit gives larger errors for the 
BH mass with respect to the one-parameter fit, we stick with the latter.
The extrapolation to smaller values of $\eta$ is consistent 
with the values obtained in Ref.~\cite{berti07} using NR simulations, also 
listed in Table~\ref{tab1}, and in Ref.~\cite{DNs} using 
a combination of test-mass limit predictions and the EOB approach. 
Henceforth, when computing the frequency and decay time of the QNMs 
we use the results obtained from the one-parameter fit in Table~\ref{tab1}.

This completes the specification of our waveform model, which can be 
applied to provide full waveform predictions for non-eccentric and  
non-spinning binary BH mergers of arbitrary mass-ratio.

\section{Comparison with numerical relativity}
\label{secnew}

\begin{figure*}
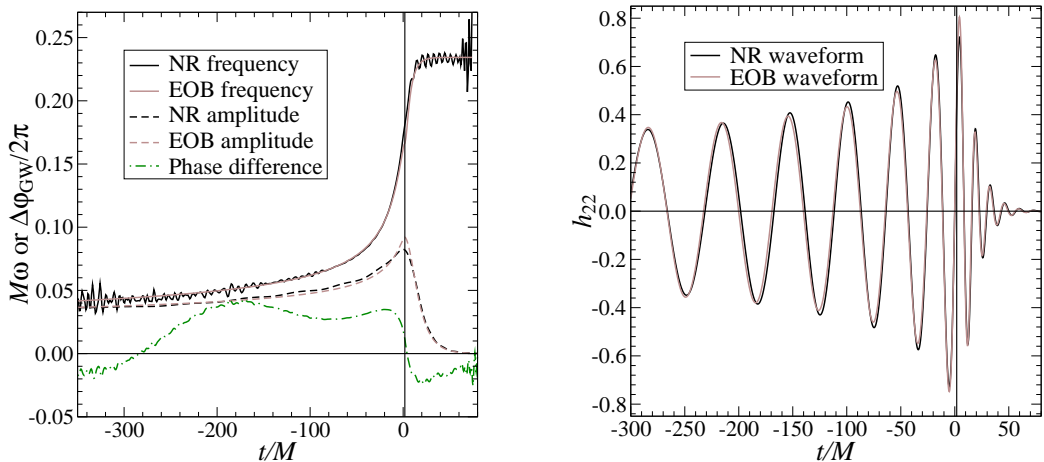

\begin{center}
\includegraphics[width=0.35\textwidth]{others8020.eps}  \hspace{1cm}
\includegraphics[width=0.35\textwidth]{waves8020.eps} 
\caption{Binary with mass ratio $4:1$. 
In the left panel we plot the NR and p4PN-EOB frequencies 
and amplitudes, and the phase difference between the EOB and NR $h_{22}$. 
In the right panel we compare the EOB and NR $h_{22}$. 
We maximize only over the initial phase and time of arrival. 
Note that we show only the last few cycles. The complete inspiral waveform
has 9 GW cycles. 
\label{fig3}}
\end{center}
\end{figure*}
In this section we examine how closely our model waveforms fit with the 
results of our numerical simulations.  While we have already applied 
some limited information from these numerical simulations in defining 
our model, such as in deriving our functional fits for $M_f(\eta)$ and 
$a_f(\eta)$ and in selecting the optimal value for $\lambda$, we can 
now compare the full waveforms.  In particular, though our model was
developed primarily in consideration of the $h_{22}$ waveforms, we find here 
that other multipolar waveform components are also well described. 

To measure the differences between the NR and EOB waveforms 
we compute the {\it fitting factor} (FF), or 
ambiguity function~\cite{DA,DIS,bcv1}. We recall that 
the overlap $\langle h_1(t),h_2(t)\rangle$ 
between the waveforms $h_1(t)$ and $h_2(t)$ is 
defined by: 
\beq
\langle h_1(t),h_2(t)\rangle\equiv4\,{\rm Re}
\int_0^\infty\frac{\tilde{h}_1(f)\tilde{h}^*_2(f)}{S_h(f)}df\,, 
\eeq
where $\tilde{h}_i(f)$ is the Fourier transform of $h_i(t)$, and
$S_h(f)$ is the detector's power spectral density (PSD). 
The FF is the normalized overlap between the NR 
waveform $h^{\rm NR}(t)$ (target) and the EOB waveform $h^{\rm EOB}(t_0,\varphi_0)$ (template) 
maximized {\it only} over the initial time $t_0$ and 
initial phase $\varphi_0$, and minimized over the initial phase 
$\varphi$ of the target (the so-called minmax~\cite{DIS}), that is 
\begin{widetext}
\beq
{\rm FF} \equiv \min_{\varphi} \max_{t_0,\varphi_0}
\frac{\langle h^{\rm NR}(\varphi;\lambda^i),h^{\rm EOB}(t_0, \varphi_0; \lambda^i)\rangle}
{\sqrt{\langle h^{\rm NR}(\varphi;\lambda^i),h^{\rm NR}(\varphi;\lambda^i) \rangle\langle 
h^{\rm EOB}(t_0, \varphi_0; \lambda^i),h^{\rm EOB}(t_0, \varphi_0; \lambda^i)\rangle}}\,,
\label{FF}
\eeq
\end{widetext}
where $\lambda^i$ are the binary parameters. For the detector PSD we shall consider either white noise 
or the LIGO noise. 

The equal-mass run lasts for $\sim 14$ GW cycles before merger. This run was published 
originally in Refs.~\cite{Goddshort,Goddlong}, and further studied for data-analysis 
purposes in Ref.~\cite{DA}. The unequal-mass runs, $m_1/m_2 = 3/2, 2, 4$, last for 
$5$, $5$, and $9$ GW cycles before merger. The $m_1/m_2 = 3/2, 2$ cases were published in 
Refs.~\cite{DA, Anatomy} and the $m_1/m_2 =4$ has recently 
been computed by the NASA-Goddard group. Adaptive mesh refinement was employed for this case as in the previous 
simulations, with a finest mesh resolution of $h_m=3M/160$ used in one run and $h_h=M/64$ 
used in a second run.  Adequate convergence of the Hamiltonian and momentum constraints were 
found; the numerical details will be reported in a future publication.  
Based on the comparisons between high- and medium-resolution waveforms,
we estimated in Ref.~\cite{DA} the FFs between high resolution and exact
waveforms for the $m_1/m_2=1$ case. Here we apply the same procedure 
for the $m_1/m_2=4$ case.
If we have several simulations with different resolutions,
specified by the mesh-spacings $x_i$, and $x_i$ are sufficiently
small, we can assume that the waveforms $h_i$ are given by
\beq 
h_i=h_0+x_i^n h_d\,, 
\eeq 
where $n$ is the convergence factor of
the waveform, $h_0$ is the exact waveform generated from the infinite
resolution run ($x_0\rightarrow0$), and $h_d$ is the leading 
order truncation error contribution to the waveform and is independent
of the mesh spacing $x_i$. The mismatch between the waveforms $h_i$ and $h_j$, 
$1-{\rm FF}_{ij}$, then scales as
\beq
1-{\rm FF}_{ij}\propto (x_i^n-x_j^n)^2\,. 
\eeq 
In the Goddard simulations, the high and medium resolution runs have mesh-spacing
ratio $x_h/x_m=5/6$, and the waveform convergence rate was shown to be $n=4$ in the
1:1 case~\cite{Goddlong}.
The FF between the high resolution and exact waveforms $h_h$ and $h_0$ is given by 
\beq 
{\rm FF}_{0h}=1-0.87 (1-{\rm FF}_{hm})\,, 
\eeq 
where FF$_{hm}$ is the FF between the high
and medium resolution waveforms $h_h$ and $h_m$. Thus, the 
mismatch between $h_h$ and $h_0$ is slightly smaller than that between
$h_h$ and $h_m$, where the latter can be derived from the FFs shown in 
the left panel of Fig.~\ref{fig4} computed using LIGO's PSD. 
If we use white noise we find FF $= 0.9922$ and $=0.9920$ 
for mass ratios 1:1 and 4:1,  respectively.
Henceforth, we shall always use high-resolution 
waveforms for the $m_1/m_2=4$ and $1$ 
cases. 

For different mass ratios, we show in the right panel of Fig.~\ref{fig4} how the FF computed 
for the dominant modes, using white noise, depends on the parameter 
$\lambda$. Based on this plot we identified 
$\lambda = 60$ as the best model, which we use in the rest of the paper. 

The p4PN-EOB model has better matching performances at the transition 
inspiral/merger/ringdown. However, the introduction of the 4PN order 
term inevitably affects the inspiral waveform. To understand the differences 
between the p4PN-EOB model and NR and other PN models  
during the long {\it inspiral phase} we plot in Fig.~\ref{figfreq} 
the frequency difference between several PN-approximants.
At the time this paper is written, preliminary results from Caltech/Cornell group
suggest that the 3PN-Tt3 approximant model fits well 
with an accurate $3000M$-long equal-mass simulation~\cite{CC}, 
we use this as the fiducial PN-model. 
The 3.5PN-T, 3PN-Tt3 and 3.5Tt1 models are 
the so-called Taylor-expanded PN models, widely used in 
the data-analysis literature (see e.g., Ref.~\cite{DISDA,bcv1,DA}).  
The plots are obtained by imposing that the PN frequencies agree 
at $\omega\,M = 0.017$. The 3PN-Tt3 model is an analytic model 
that uses the PN-expanded phase as function of time. It 
was also used in Ref.~\cite{BCP} [see Eq.~(31) there], where it 
was found to best-match (together with the 3.5PN-T model) 
the equal-mass waveform computed with generalized harmonic 
coordinates by Pretorius. The 3.5PN-T model uses 
the energy balance equation re-expanded in powers of the orbital 
frequency. It was found to best-match the equal-mass waveforms 
in Refs.~\cite{BCP,Goddshort,DA}. The 3.5PN-Tt1 model is a numerical PN model 
that solves the energy balance equation without re-expanding 
the flux and energy function, as done in the 3.5PN-T model. 
From the plots we conclude that while for the equal-mass case 
the p4PN-EOB model is rather close to our fiducial model (and thus to 
preliminary results from long numerical simulations), 
we cannot draw definite conclusions 
for generic mass ratios. Indeed, the right panel in Fig.~\ref{figfreq} shows  
that for unequal masses, the closeness of the PN-approximants is different 
than in the equal-mass case. 

\begin{figure*}
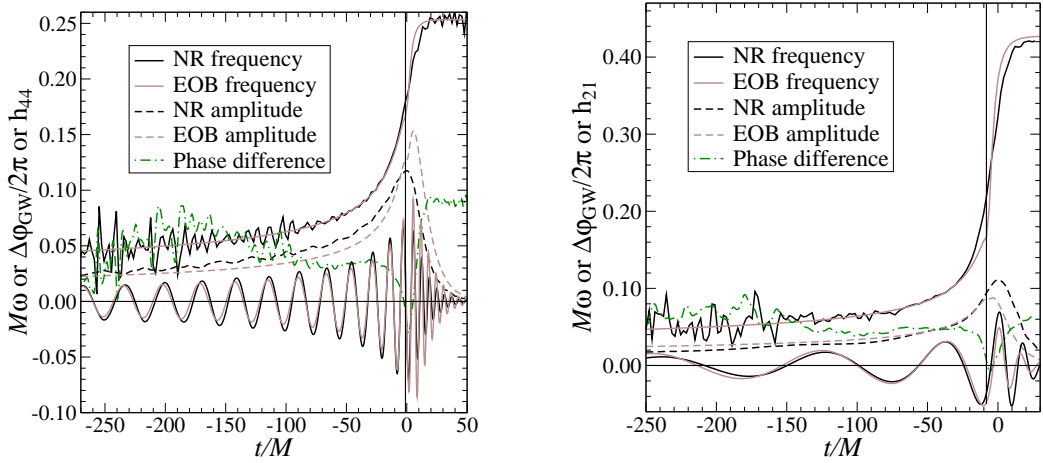

\begin{center}
\includegraphics[width=0.35\textwidth]{C44waves4010.eps} \hspace{1cm}
\includegraphics[width=0.35\textwidth]{C21waves4010.eps}  
\caption{Binary with mass ratio $4:1$. In the left (right) panel we compare the p4PN-EOB and NR 
$h_{44}$ ($h_{21}$). We maximize only on the initial phase and time of arrival. \label{fig6}}
\end{center}
\end{figure*}

In Figs.~\ref{fig2}, \ref{fig3}, and \ref{fig6} we show the 
comparison between the p4PN-EOB and NR waveforms, orbital frequencies, and 
phase differences for the most accurate, long numerical simulations, notably 
the 1:1 and 4:1 cases. We show results for the $l=2,m=2$ mode, and 
also for the $l=4,m=4$ and $l=2, m=1$ modes. The p4PN-EOB $l=3,m=3$ mode matches 
rather well the NR mode, similarly to the $l=2,m=2$ case, thus we do not show it. 
When the mass-ratio increases the other modes are no-longer 
so subdominant with respect to the $l=2, m=2$ mode, 
as seen in the right panel of Fig.~\ref{fig5}. 
In Appendix~\ref{appendix_massratio} we present 
similar plots for the $l=2,m=2$ mode of the shorter runs with 
mass ratios 2:1 and 3:2. 

\begin{table}[t]
\begin{center}
{\begin{tabular}{|l|c|c|c|}
\hline $\eta$ & ${l m}$ &$\langle h^{\rm NR}_{lm}, h^{\rm p4PN-EOB}_{lm} \rangle $&
$\Delta \varphi_{\rm GW}/(2 \pi)$ \\ \hline\hline 
$0.25$ & ${22}$ & 0.9907 & $\pm 0.030$ \\\hline
$0.24$ & ${22}$ & 0.9881 & $\pm 0.058$ \\\hline
$0.22$ & ${22}$ & 0.9878 & $\pm 0.078$ \\\hline
$0.16$ & ${22}$ & 0.9925 & $\pm 0.035$ \\
$0.16$ & ${33}$ & 0.9860 & $\pm 0.055$ \\
$0.16$ & ${44}$ & 0.9436 & $\pm 0.065$ \\
$0.16$ & ${21}$ & 0.9092 & $\pm 0.050$ \\
\hline
\end{tabular}}\end{center}
\caption{For several mass configurations, we list the (minmax) FF 
obtained using white noise and maximizing only on the 
time of arrival and initial phase, and the phase difference 
in one GW cycles. For comparison, using 
the 3.5PN-EOB model we find $\langle h^{\rm NR}_{22}, h^{\rm 3.5PN-EOB}_{22} \rangle  = 
0.8718$ and = $0.9569$ for mass ratios 1 and 4, respectively.
\label{tab2}}
\end{table}

In Tables~\ref{tab2}, \ref{tab3} and \ref{tab4} we list the FFs 
and the phase difference (in one GW cycle) between several p4PN-EOB 
modes and the NR modes, for white noise and LIGO's PSD, respectively.
The dominant frequencies associated with each $l$, $m$ mode are rather similar 
all along the inspiral, as seen in the left panel of 
Fig.~\ref{fig5}. However, due to the different frequency of the 
fundamental QNM~\cite{Anatomy}, the frequencies associated with 
the $l=2$, $m=1$ and $l=3$, $m=2$ modes decouple 
from the other frequencies during the transition inpiral(-plunge) to ringdown. 

Tables~\ref{tab2}, \ref{tab3} show that the FFs are rather 
high except for a few modes, like the $l=4$, $m=4$ mode 
with mass-ratio 4:1 [see the left panel in Fig.~\ref{fig6}]. 
In this case we find that our matching procedure is not 
so efficient in reproducing the amplitude of the NR ringdown waveform. 
More studies extended to different mass ratios may shed 
light on this anomaly. The $l=3$, $m=2$ mode contains a mode mixing 
between the $l=2$, $m=2$ and $l=3$, $m=2$ modes~\cite{BCP}, and 
the matching procedure that we adopt does not accurately reproduce it. 
Better knowledge of how these modes are excited during the inspiral to 
ringdown transition is required to solve this problem.
Finally, because the dominant frequency 
associated with the $l=2$, $m=1$ mode departs from the 
dominant frequency associated with the $l=2$, $m=2$ mode quite 
before the merger (see the left panel in Fig.~\ref{fig5}), 
and it rises to a much higher QNM frequency, we find 
that the FF associated with the $l=2$, $m=1$ mode is not 
very high. In fact, our matching procedure is optimized 
to reproduce the increase of frequency of the dominant 
$l=2$, $m=2$ and $l=3$, $m=3$ modes. 

While the FFs listed in Tables~\ref{tab2}, \ref{tab3} are obtained maximizing 
{\it independently} over the phase and time of arrival of  
each $l$, $m$ mode, the FFs in Table~\ref{tab4} are computed by matching the 
full waveform containing all-together the leading modes (see Fig.~\ref{fig5}).  
To achieve this we build the numerical and EOB expressions for
\begin{widetext}
\bea
h(\theta,\phi; t_0, \varphi_0;\lambda_i) &\equiv& -(h_+ - i h_\times)\,, \nonumber \\
&&= {}_{-2}Y_{21}(\theta,\phi)\, h_{21}(t_0, \varphi_0;\lambda_i) + {}_{-2}Y_{22}(\theta,\phi)\,h_{22}(t_0, \varphi_0;\lambda_i)
+ {}_{-2}Y_{31}(\theta,\phi)\,h_{31}(t_0, \varphi_0;\lambda_i) + \nonumber \\
&& {}_{-2}Y_{32}(\theta,\phi)\,h_{32}(t_0, \varphi_0;\lambda_i)+ 
{}_{-2}Y_{33}(\theta,\phi)\,h_{33}(t_0, \varphi_0;\lambda_i)+ {}_{-2}Y_{42}(\theta,\phi)\,h_{42}(t_0, \varphi_0;\lambda_i) +
\nonumber \\
&&  {}_{-2}Y_{44}(\theta,\phi)\,h_{44}(t_0, \varphi_0;\lambda_i) +
{}_{-2}Y_{2-1}(\theta,\phi)\, h_{2-1}(t_0, \varphi_0;\lambda_i) + {}_{-2}Y_{2-2}(\theta,\phi)\,h_{2-2}(t_0, \varphi_0;\lambda_i)
+ \nonumber \\
&&  {}_{-2}Y_{3-1}(\theta,\phi)\,h_{3-1}(t_0, \varphi_0;\lambda_i) + 
{}_{-2}Y_{3-2}(\theta,\phi)\,h_{3-2}(t_0, \varphi_0;\lambda_i)+ 
{}_{-2}Y_{3-3}(\theta,\phi)\,h_{3-3}(t_0, \varphi_0;\lambda_i) +\nonumber \\
&&  {}_{-2}Y_{4-2}(\theta,\phi)\,h_{4-2}(t_0, \varphi_0;\lambda_i) + {}_{-2}Y_{4-4}(\theta,\phi)\,h_{4-4}(t_0, \varphi_0;\lambda_i)\,, 
\label{h}
\eea
\end{widetext}
and maximize over the template time-of-arrival $t_0$ and initial phase
$\varphi_0$. The waveform seen by an interferometer GW detector is 
the linear combination of the two polarization states
$h_+$ and $h_\times$ defined in the radiation frame, with the
combination coefficients given by the polarization angle $\alpha$, as
$\cos\alpha$ and $\sin\alpha$ [see e.g., Ref. ~\cite{MF}]. 
Since in Eq.~\eqref{h} the $h_{lm}$ modes have different dependence on 
$\varphi_0$, the polarization angle $\alpha$ can not be absorbed 
into $\varphi_0$, as done when considering only the dominant 
$l=2, m=2$ mode. In principle, we should consider (target) 
numerical signals with different values of $\alpha$ and maximize 
the EOB template over $\alpha$. However, since we shall find 
that both $h_+$ and $h_\times$ can be matched with high FFs 
by the p4PN-EOB templates, it is not likely that different 
values of $\alpha$ will reduce the FFs significantly. 

\begin{table}[t]
\begin{center}
{
\begin{tabular}{|l|c|c|c|}
\hline $(m_1+m_2)$ & ${l m}$ &$\langle h^{\rm NR}_{lm}, h^{\rm p4PN-EOB}_{lm} \rangle $ &
$\Delta \varphi_{\rm GW}/(2 \pi)$ \\ \hline\hline 
$(15+15) M_\odot$ & ${22}$ & 0.9975 & $\pm 0.043$ \\
$(50+50) M_\odot$ & ${22}$ & 0.9817 & $\pm 0.043$ \\\hline
$(20+30) M_\odot$ & ${22}$ & 0.9897 & $\pm 0.065$ \\\hline
$(20+40) M_\odot$ & ${22}$ & 0.9889 & $\pm 0.068$ \\\hline
$(10+40) M_\odot$ & ${22}$ & 0.9961 & $\pm 0.035$ \\
$(10+40) M_\odot$ & ${33}$ & 0.9911 & $\pm 0.055$ \\
$(10+40) M_\odot$ & ${44}$ & 0.9720 & $\pm 0.075$ \\
$(10+40) M_\odot$ & ${21}$ & 0.9737 & $\pm 0.080$ \\\hline
$(20+80) M_\odot$ & ${22}$ & 0.9965 & $\pm 0.035$ \\
$(20+80) M_\odot$ & ${33}$ & 0.9873 & $\pm 0.055$ \\
$(20+80) M_\odot$ & ${44}$ & 0.9548 & $\pm 0.065$ \\
$(20+80) M_\odot$ & ${21}$ & 0.9804 & $\pm 0.125$ \\
\hline
\end{tabular}}\end{center}
\caption{For several mass configurations, we list the (minmax) FFs 
obtained using LIGO's PSD and maximizing only on the 
time of arrival and initial phase, and the phase difference 
in one GW cycles. 
\label{tab3}}
\end{table}

\begin{figure}
\begin{center}
\includegraphics[width=0.35\textwidth]{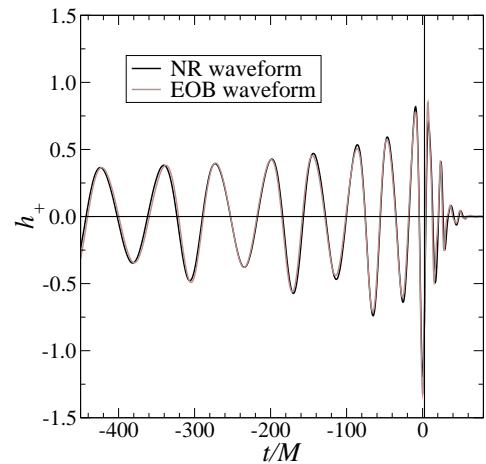}
\caption{Binary with mass ratio $4:1$. We compare the p4PN-EOB and NR 
$h_+$, given by Eq.~(\ref{h}), extracted at inclination angle 
$\theta = \pi/3$, maximizing only over the initial phase and time of arrival. 
\label{fig7}}
\end{center}
\end{figure}

We notice that to reproduce $h_+$ and $h_\times$ through 1PN order
as given by Ref.~\cite{BIWW}, we would need to add the next-to-leading
PN correction to $h_{22}$ in Eq.~(\ref{22}), that is we need to
multiply Eq.~(\ref{22}) by $[1 + (55 \eta - 107)/42\,\omega^{2/3}]$,
and insert Eqs.~(\ref{22})--(\ref{44}) in Eq.~(\ref{h}). After
checking that the next-to-leading correction to $h_{22}$ has a tiny
effect on the FF, and that also the contribution of the modes
$h_{32}$, $h_{31}$ and $h_{42}$ is rather small, we compute the FFs
restricting Eq.~(\ref{h}) to the dominant modes $h_{22}$, $h_{33}$,
$h_{44}$, and $h_{21}$. The figures listed in Table~\ref{tab4} refer
to different inclination angles.  Despite the fact that the $l=4$,
$m=4$ mode is recovered only with FF$ \sim 0.95$ and the $l=2$, $m=1$
mode is recovered only with FF$ \sim 0.90$ (see Tables~\ref{tab2} and
\ref{tab3}), the FFs obtained by combining all the modes are rather
high. This happens because the amplitude of the $l=4$, $m=4$ and
$l=2$, $m=1$ modes are much lower than the amplitude of the $l=2$,
$m=2$ and $l=3$, $m=3$ modes, as seen in Fig.~\ref{fig5}. 
In Fig.~\ref{fig7} we plot the 4PN-EOB and NR $h_+$ waveforms for 
the case of inclination angle $\theta = \pi/3$. 

The  waveforms shown in Figs.~\ref{fig4}, \ref{fig2}, \ref{fig3}, \ref{fig6} and 
\ref{fig7}, are {\it normalized} waveforms. In the right panel of 
Fig.~\ref{fig5}, we compare the NR and EOB restricted-approximated amplitudes for 
several $l$ and $m$ modes. By restricted-approximated amplitude we mean 
that for each $l$ and $m$ we restrict ourselves to the leading order 
term in the PN expansion [see Eqs.~(\ref{22})--(\ref{44})]. 
We notice, as already pointed out in Ref.~\cite{DA}, 
a non-negligible difference between the NR and EOB amplitude.  
Higher-order PN corrections in the amplitude could in part nullify the difference~\cite{Goddlong,NewJena}, 
but due to the oscillatory behaviour of the higher PN corrections~\cite{BCP}, 
it is hard to draw a robust conclusion. Moreover, there could 
be a systematic error in the numerical amplitude due to 
extraction radius. From Fig.~\ref{fig5} we see that 
amplitudes computed at different resolutions do not affect 
the amplitude significantly. Although, the difference  
in the amplitudes has negligible effects on the FFs~\cite{DA}, 
it may affect the binary parameter estimation and 
the prediction of the recoil velocity from merging binary BHs~\cite{Anatomy}, 
thus it needs to be sorted out.

The difference in amplitudes also affects the energy and angular-momentum 
released during the ringdown. We could scale the EOB amplitude
by a constant factor and get the same energy and angular momentum as in the 
NR simulation at the time the ringdown waveform is attached. However, we find 
that the amount of mass and angular momentum released during the ringdown 
stage is always larger than the values predicted by NR, as can be seen from the
amplitudes shown in the left panels of Figs.~\ref{fig2} and
\ref{fig3}. Therefore, the mass and spin of the final BH computed from
the EOB model itself are always smaller than the values predicted 
by NR. For example considering the $h_{22}$ mode, we find that the 
EOB model underestimate the final mass and spin of the BH
by $\sim 2\%$ and $\sim 8\%$. To resolve this inconsistency, 
the EOB amplitudes in both the inspiral and ringdown phased 
would need to be improved. Consequently, we also notice that if we used a 
{\it bootstrap} mechanism to locate the 
EOB matching point between inspiral and ringdown, we would find that 
the EOB matching must occur too early in time to compensate the fact that the 
EOB model overestimates the energy and angular momentum released 
during the ringdown. Such an early matching point would cause a significant 
phase difference between the EOB and NR waveforms.  

\section{Conclusions}
\label{sec5}

In this paper we have begun to exploit the flexibility of the 
EOB approach~\cite{TD} to analytically encode information 
from numerical relativity simulations, in order to build faithful templates for binary BHs. 
By faithful templates~\cite{DIS} we mean templates that have
large overlaps, say $ \gaq \, 96.5\%$, with the 
expected signal achieved by maximizing {\it only} over the 
initial phase and time of arrival. Thus,  they can be 
employed for detection and parameter estimation.

We have built accurate waveforms by improving the EOB model 
as follows: We add a p4PN-order term $\lambda \eta$
($\lambda = 60$) to the EOB coefficient $A(r)$, and 
Pad\'e-approximate the latter to guarantee the presence 
of the LSO and the light-ring~\cite{DJS}. Note that 
the inclusion of a 4PN term in the coefficient $A(r)$ 
was employed in Ref.~\cite{DGG} to better fit the EOB model 
to quasi-equilibrium initial-data configurations~\cite{GGB}.  
The choice $\lambda = 60$ found in this paper provides  
better agreement between the EOB and NR GW frequency 
associated with the $l=2$, $m=2$ mode during 
the transition inspiral(-plunge) to ringdown. 
The complete waveform, given by Eqs.~(\ref{22})--(\ref{44}) and Eq.~(\ref{h}),  
is then built by evolving Eqs.~(\ref{eq:eobhamone})--(\ref{eq:eobhamfour}) 
throughout the inspiral(-plunge) and attaching three QNMs at the peak of the 
p4PN-EOB GW frequency, as determined by Eq.~(\ref{omegamatch}). 
The QNM frequency and decay time are fixed by the final BH mass and spin, which 
can be predicted by extrapolating sparse NR results through a fit which meets expectations for the test-mass limit (see Table~\ref{tab1}). 
This is distinct from other approaches based on the EOB model itself~\cite{DNs}. 
Our general approach remains flexible, allowing the possibility of future improvements.

\begin{table}[t]
\begin{center}
{
\begin{tabular}{|l|c|c|}
\hline $\theta$ & $\langle h_+^{\rm NR}, h_+^{\rm p4PN-EOB} \rangle $ & 
$\langle h_\times^{\rm NR}, h_\times^{\rm p4PN-EOB} \rangle $ \\ \hline\hline 
$0 $ &  0.9925 & 0.9925 \\
$\pi/6$ & 0.9915 & 0.9917\\
$\pi/4$ & 0.9890 & 0.9900\\
$\pi/3$ & 0.9854 &  0.9883\\
$\pi/2$ & 0.9803 & $-$ \\\hline
\end{tabular}}\end{center}
\caption{Binary with mass-ratio 4:1. 
For several inclinations angles, we list the (minmax) FFs 
between the p4PN-EOB and NR waveforms, given by Eq.~(\ref{h}), 
using white noise. In all cases the fitting factor exceeds $0.98$.
\label{tab4}}
\end{table}
\begin{figure*}
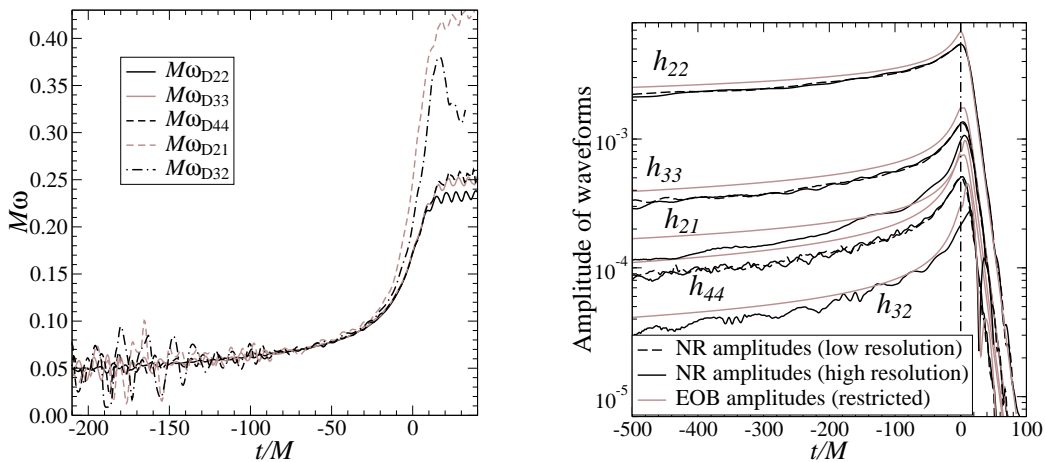

\includegraphics[width=0.35\textwidth]{freqs41.eps} \hspace{1cm}
\includegraphics[width=0.35\textwidth]{amps41.eps} 
\caption{Binary with mass ratio $4:1$. In the left panel we plot the 
dominant frequencies for several $l$ and $m$ modes. Generalizing 
Eq.~(23) in Ref.~\cite{BCP}, we define $\omega_{Dlm} = -(1/m)\, {\rm Im}({}_{-2}\dot{C}_{lm}/{}_{-2}C_{lm})$ 
~\cite{Anatomy}. In the right panel we compare the NR amplitude and EOB 
restricted-approximated amplitude for 
several $l$ and $m$ modes. By restricted-approximated amplitude we mean 
that for each $l$ and $m$ we restrict ourselves to the leading order term 
as given by Eqs.~(\ref{22})\mbox{--}(\ref{44}).  
\label{fig5}}
\end{figure*}

Although we have tested the accuracy of the p4PN-EOB model for 
mass ratios $\eta = 0.25, 0.24, 0.22$, and $0.16$, 
our procedure also predicts the waveforms for values of $\eta < 0.16$, 
that can be tested against numerical simulations when 
accurate, long waveforms become available. 
Currently, for all cases considered we obtain phase differences 
with the NR waveforms of less than $ \pm 8\%$ in a GW cycle. 
This result refers to waveforms containing at most $14$ GW cycles.  
To test the perfomances of the p4PN-EOB model during the long 
inspiral phase, we compared this model with other PN-approximants which 
fit rather well extremely accurate NR simulations 
(see Fig.~\ref{figfreq}). Before merger, in the equal-mass case, 
the p4PN-EOB model is rather close to the 3PN-Tt3 model which
may agree well with long-lasting numerical simulations~\cite{CC}.
But we cannot draw definite conclusions about our model's inspiral performance 
because the results depend on the mass ratios, and extremely accurate,  
long-lasting NR simulations are not available, yet, for unequal masses. 
Therefore, we do not exclude other possible adjustments in the 
EOB model to keep track of the phase 
evolution for an extremely large number of GW cycles. 

Using white noise and LIGO's PSD, we found FFs $\gaq\, 0.98$, except 
for the $l=4$, $m=4$, $l=3$, $m=2$ and $l=2$, $m=1$ modes of mass ratio 4:1 
(see Table~\ref{tab2}, ~\ref{tab3}). The matching procedure from inspiral(-plunge) 
to ringdown that we employ is too simplified when QNMs with different 
$l$ are present and/or the dominant frequency associated 
with the $l$, $m$ mode is rather different from the 
dominant frequency associated with the $l=2$, $m=2$ mode. 
This can show up as an earlier decoupling and higher 
QNM frequency (see the left panel of Fig.~\ref{fig5}). 
We shall explore a more suitable matching procedure in 
the near future. In any case, when building the 
complete waveform containing all the relevant $l$, $m$ modes 
[see Eq.~(\ref{h})] and takes into account their different 
amplitudes (see Fig.~\ref{fig5}), we obtain rather high FFs, 
as seen in Table~\ref{tab4}.

We pointed out, as already done in Ref.~\cite{DA}, 
that there exists a non-negligible difference in the amplitude 
of the PN (EOB or Taylor-expanded-PN) and NR waveforms, whose 
origin has not yet been accounted for (see the right panel in Fig.~\ref{fig5}). 
It might be due to higher-order PN corrections in the 
amplitude~\cite{BCP,Goddlong,NewJena}. 

When maximizing over the binary masses, the p4PN-EOB template 
family presented here will have extremely high matching performances 
and can be used for coherent detection of non-spinning 
binary BHs, further improving the EOB model presented in Ref.~\cite{DA}. 
Once the amplitude difference is resolved, 
the p4PN-EOB templates can be employed 
for parameter estimation for ground-based detectors.
For LISA, which is expected to 
observe the coalescence of supermassive 
black holes with rather high signal-to-noise 
ratios, say $\sim 10^2\mbox{--} 10^4$,
the faithfulness to the model presented here is not 
high enough to allow accurate parameter estimation.  
Improvements to the EOB model, tuned to match 
the results of longer-lasting, and more highly accurate
numerical simulations will be necessary.
In this respect we notice that important advances 
have been made in solving numerically 
the two-body problem in the test-mass-limit case using Regge-Wheeler-Zerilli-type methods, 
and comparing it to the EOB model~\cite{DNT,DNf}. The authors of Refs.~\cite{DNT,DNf} 
introduced several improvements in the EOB approach. Notably, they  
include non-adiabatic quasi-circular orbit terms, match the 
inspiral(-plunge) waveform to the ringdown modes over a time-interval, 
instead of at one point, and include up to five ringdown overtones. These 
refinements lead to a reduction of the phase difference between 
the numerical and EOB waveforms from $\pm 3.5\%$ to $\pm 1.1 \%$, and  
more importantly, to excellent agreement of the $l=2$, $m=2$ mode amplitude. 
Soon these improvements will be applied also in the comparable-mass case. 

The results presented here point to the possibility that an essentially complete understanding 
of binary BH mergers, developed with the aid of 
highly-accurate NR simulations, may be encoded in 
relatively simple, convenient analytic models for GW 
data analysis applications, avoiding the need for 
templates derived directly from NR simulation.
In the future we plan to improve this analysis by 
including higher-order PN amplitude corrections and to explore extensions
of the model to spinning, precessing binary systems, using improved versions 
of the EOB model with spin~\cite{TD,BCD}. 

\acknowledgments
We thank Jeremy Schnittman for useful comments and interactions, and 
Greg Cook for an informative discussion. 
A.B. and Y.P. acknowledge support from NSF grant PHY-0603762, and A.B. also acknowledges support
from the Alfred P Sloan Foundation. 
The work at Goddard was supported in part by NASA grant
06-BEFS06-19.  The simulations were carried out using
Project Columbia at the NASA Advanced Supercomputing Division (Ames
Research Center) and at
the NASA Center for Computational Sciences (Goddard
Space Flight Center).  B.K. was supported by the NASA Postdoctoral 
Program at the Oak Ridge Associated Universities.
S.T.M. was supported in part by the Leon A. Herreid Graduate Fellowship.

\appendix

\section{Comparison with binaries of mass ratio 2:1 and 3:2}
\label{appendix_massratio}

\begin{figure*}
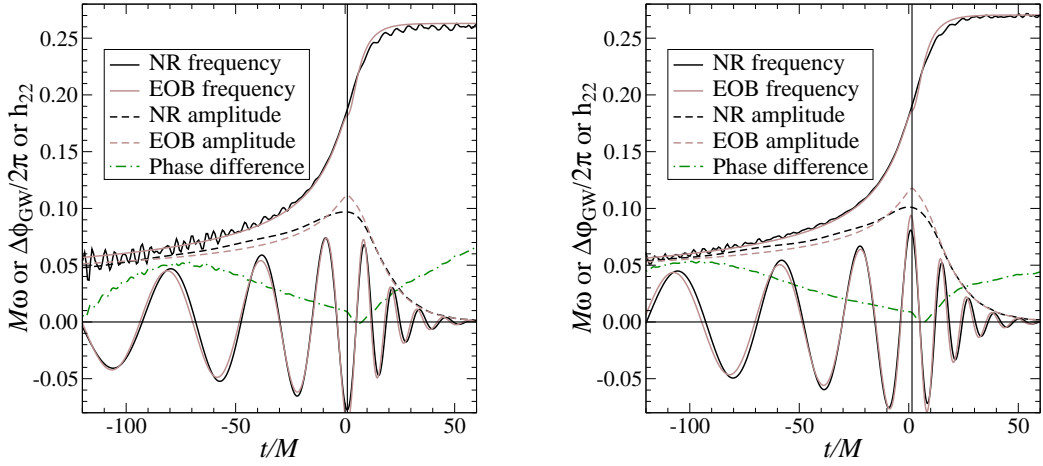

\begin{center}
\includegraphics[width=0.35\textwidth]{waves4020.eps}  \hspace{1cm}
\includegraphics[width=0.35\textwidth]{waves3020.eps} 
\caption{In the left (right) panel we plot the NR and p4PN-EOB frequencies 
and amplitudes, and the phase difference between the p4PN-EOB and NR $h_{22}$ 
for a binary with mass ratio 2:1 (3:2).
We maximize only over the initial phase and time of arrival.  \label{fig8}}
\end{center}
\end{figure*}

In Fig.~\ref{fig8} we show how the p4PN-EOB model performs for mass ratios
2:1 and 3:2. In these cases the waveforms are much shorter, and 
the eccentricity during the inspiral case is more pronounced. 
Details on these simulations can be found in Refs.~\cite{DA, Anatomy}.

\end{document}